\documentclass[twocolumn,nobibnotes,showpacs,preprintnumbers,amsmath,amssymb]{revtex4}
\pdfoutput=1
\usepackage{graphicx, hyperref}% Include figure files
\usepackage{dcolumn}% Align table columns on decimal point
\usepackage{amsmath}%include nice math fonts, etc.
\usepackage{bm}% bold math

\newcommand{\be}{\begin{equation}}
\newcommand{\ee}{\end{equation}}
\newcommand{\bea}{\begin{eqnarray}}
\newcommand{\eea}{\end{eqnarray}}

\newcommand{\MeV}{\,{\rm MeV}}
\newcommand{\GeV}{\,{\rm GeV}}
\newcommand{\TeV}{\,{\rm TeV}}

\newcommand{\eqnref}[1]{Eq.~(\ref{#1})}
\newcommand{\figref}[1]{Fig.~\ref{#1}}

\begin{document}
\title{Correlated Signals at the Energy and Intensity Frontiers from Nonabelian Kinetic Mixing}
\author{G.~Barello, Spencer Chang and Christopher A.~Newby}

\affiliation{Department of Physics and Institute of Theoretical Science, University of Oregon, Eugene, Oregon 97403}

\begin{abstract}
We show that when a dark abelian gauge sector and SU$(2)_{L}$ kinetically mix it necessarily generates a relation between the kinetic mixing strength and the mass of the mediating particle. Remarkably, this correspondence maps the weak scale directly to the kinetic mixing strengths being probed by modern fixed-target experiments and next generation flavor factories. This illuminates the exciting possibility of correlated discoveries of a new particle at the LHC and a dark photon at intensity frontier experiments. To motivate the scenario, we construct a simple model and explore its phenomenology and constraints.
\end{abstract}

\maketitle

Kinetic mixing (KM) is a phenomenon that produces an interaction between gauge bosons of two different gauge groups, and generically occurs when there are two U$(1)$ gauge symmetries in a theory. KM will be generated by loop processes whenever there are particles charged under both symmetries \cite{Galison:1983pa,Holdom:1985ag}.  This makes KM a common ingredient of models beyond the standard model (SM) of particle physics. In $Z'$ \cite{Galison:1983pa} and many dark matter models \cite{Pospelov:2007mp, ArkaniHamed:2008qn}, the SM is supplemented by an additional U$(1)$ gauge symmetry which can mix with U$(1)_{Y}$.  Such models have motivated a large and diverse experimental effort with current and upcoming searches at intensity frontier experiments (fixed-target and flavor factories) and the LHC (see  \cite{Essig:2013lka} for overview and references). The main focus of these searches and models has been on the dynamics of the dark photon or signals of particles charged only under the dark sector, while little attention has been paid to the aforementioned particle charged under both symmetries, which mediates KM.\\

The reason this mediating particle is ignored is that it's mass can usually be made large while leaving the KM strength fixed.  In the most studied case of KM between two abelian sectors, where the mixing operator is dimension four, the mediator mass only logarithmically affects the strength of KM. Moreover, KM between abelian sectors is described by a renormalizable operator, so it can be included without reference to a mediator.  Thus, in the abelian case, it is not guaranteed that the mediator will be light enough to be discovered.  On the other hand, when KM goes through a nonabelian gauge sector, the operator is nonrenormalizable and inextricably linked to a mass scale.  This fact gives nonabelian kinetic mixing models unique predictive power which has not yet been studied in the literature. This study fills that gap. Furthermore, as we will show, nonabelian KM strengths relevant for current intensity frontier experiments is unambiguously linked to a weak scale mediator, predicting a correlated signal at the energy frontier.   Although such nonabelian mixing is already well known in the literature (see, for example, \cite{ArkaniHamed:2008qn}) this study represents the first statement of this connection, and the first presentation of a model where a nonabelian operator is the sole origin of KM.

In this paper we discuss a case of particular modern interest: an abelian dark sector mixing with SU$(2)_{L}$ of the SM. The lowest dimensional operator involving only SM fields and the dark photon which kinetically mixes SU$(2)_L$ and the dark photon is
\begin{equation}
\frac{c}{16\pi^2m_\phi^{2}}\left(H^{\dag} \tau^{a} H\right) W^{a}_{\mu\nu} F_{D}^{\mu\nu}
 \label{eq:kin mix op}
\end{equation}
\noindent where $W^a_{\mu\nu} (F_{D}^{\mu\nu})$ is the field strength of the SM SU$(2)_L$ gauge boson (dark gauge boson), $H$ is the SM higgs field, and $\tau^a$ are the Pauli matrices divided by two.  Anticipating the origin of this operator, we include the mass of the mediator $m_\phi$, a loop factor, and absorb $O(1)$ numbers and couplings into the coefficient $c$.   Once electroweak symmetry is broken, \eqnref{eq:kin mix op} contains the canonical mixing between the photon and the dark photon
%%
%\begin{equation}
%\frac{\epsilon}{2} F_{\mu\nu} F_{D}^{\mu\nu}= \frac{(v^2/2\Lambda^2)s_{W}}{2}F_{\mu\nu} F_{D}^{\mu\nu}
%\label{eq:canonical mixing}
%\end{equation}
%%
%
%%
\begin{equation}
\frac{\epsilon}{2} F_{\mu\nu} F_{D}^{\mu\nu};\hspace{.3in} \epsilon =  \frac{c\,  v^2 s_{W}}{32\pi^2m_\phi^2}
\label{eq:canonical mixing}
\end{equation}
\noindent where $s_{W}$ is the sine of the electroweak mixing angle, and $v$ is the SM higgs vacuum expectation value (vev).  Already this expression shows a connection between intensity and energy frontier experiments: planned searches for the dark photon include  $i)$ fixed target experiments, probing the region $\epsilon \sim 10^{-5} - 10^{-4}$ for a dark photon of mass $M_{A_{D}} \sim 10-200 \MeV$ and $\epsilon \gtrsim 3\times 10^{-4}$ for $M_{A_{D}} \sim 10-600 \MeV$  ({\it e.g.} APEX \cite{Essig:2010xa} and HPS \cite{Celentano:2014wya}), $ii)$ next generation flavor factories,  sensitive to $\epsilon \sim 10^{-4} - 10^{-3}$ for dark photon masses up to 10 GeV \cite{Essig:2013lka} (going beyond existing BABAR, BESIII limits  \cite{Lees:2014xha, Prasad:2015mxa}), and $iii)$ a proposed LHCb search sensitive to the range $\epsilon \sim 10^{-5} - 10^{-3}$ and $M_{A_{D}} \leq 100 \MeV$ \cite{Ilten:2015hya}.   In our models of interest, \eqnref{eq:canonical mixing} shows that this parameter space requires  
\begin{equation}
m_\phi =  \sqrt{\frac{c\, v^2 s_W}{32\pi^2 \epsilon}} \sim \sqrt{\frac{c}{\epsilon/10^{-4}}} \times 1\TeV.
\label{eq:canonical mixing}
\end{equation}
%%
%In our models of interest, these experiments are sensitive to $\epsilon = (v^2/2\Lambda^2)s_{W} \sim 10^{-4}$, leading to a scale of new physics  $\Lambda \sim 12 \TeV$.  If this operator is generated at one loop order by perturbative couplings, $1/\Lambda^2 \lesssim 1/(16\pi^2 m_\phi^2)$, leading to $m_\phi \lesssim \TeV$.  
Thus, in theories with only nonabelian kinetic mixing, there is a strong correlation between signals of dark photons at the intensity frontier and the corresponding mediator particles at the LHC. This conclusion is independent of the specific realization of nonabelian KM. 

In the rest of this paper we present a simple model where the only KM that occurs is nonabelian.  In such scenarios, the mediator particle's signals at the LHC are correlated with the dark photon searches of the intensity frontier.  We will analyze the model's dynamics and then discuss the mediator particle's phenomenology and relevant constraints.

\vspace{.1cm} \noindent {\bf\emph{Model:}}  In this model, there is a dark gauge symmetry U$(1)_{D}$ with a dark photon, $A_D$. The field mediating KM is a scalar SU$(2)_{L}$ triplet with unit dark charge that we call $\phi$. In order to give the dark photon mass we introduce a dark higgs, $H_D$,  with unit dark charge that gets a vev $\langle H_{D}\rangle = v_{D}/\sqrt{2}$. The most general, renormalizable theory with these fields has many terms in its scalar potential. Only a subset of them will be relevant for our discussion, and the terms we study are
\begin{eqnarray}
V(H,H_{d},\phi) &= &\lambda |H|^{4} - \mu^{2} |H|^{2} + \lambda_{D} |H_{D}|^{4} - \mu_{D}^{2} |H_{D}|^{2} \nonumber \\
 &&+ m_{\phi}^{2} |\phi|^{2} + \lambda_{\text{mix}} (\phi^\dagger T^{a} \phi)(H^\dagger \tau^{a}H) \nonumber \\
 &&+ \kappa\left[\phi^{a}(H^\dagger \tau^{a} H)H_{D}^\dagger + \mbox{h.c.}\right] 
\label{eq:scalar potential}
\end{eqnarray}
\noindent where $\kappa$ can be taken to be real after a field redefinition and $T^a$ is the triplet representation's generators for SU$(2)_L$.  Of particular importance is the term with coefficient $\lambda_{\text{mix}}$ as it is responsible for KM.  After integrating out $\phi$, KM is generated with strength
\begin{equation}
 \epsilon = \frac{g g_D\lambda_\text{mix}}{96\pi^2}\frac{v^2}{m_\phi^2}s_W \sim 10^{-4}\, g_D\, \lambda_\text{mix} \left(\frac{400 \text{ GeV}}{m_\phi}\right)^2
\label{eq:epsilonexp}
 \end{equation}
\noindent where $g$ is the gauge coupling for SU$(2)_L$, and $g_{D}$ is the dark gauge coupling.  As the final expression shows, if the new couplings are order one, mixings relevant to intensity frontier experiments are spanned by $m_{\phi}$ in the range $100 \text{ GeV} - 1 \text{ TeV}$.

This model does not contain a particle charged under both U$(1)_{D}$ and hypercharge so there is no abelian kinetic mixing generated at low energies.  Although an ultraviolet contribution to abelian kinetic mixing can exist, it can be suppressed if the Standard Model is embedded into a grand unified theory (GUT). That way there is no abelian kinetic mixing in the ultraviolet, since hypercharge becomes part of a nonabelian group.  In such scenarios, particles with GUT-scale masses can generate abelian kinetic mixing from two-loop diagrams with mixing strengths on the order $\epsilon \sim  10^{-6} - 10^{-4}$ as discussed in \cite{ArkaniHamed:2008qp}. In this and similar models, nonabelian kinetic mixing can be dominant over abelian mixing. This means that we can use \eqnref{eq:epsilonexp} to predict the mediator mass from the kinetic mixing strength.

\vspace{.1cm} \noindent \emph{Mass Spectrum:} The term responsible for KM also generates a mass splitting in the $\phi$ states. Two states, labeled $\chi^\pm$ and $\eta^\pm$, are charged under electromagnetism and have masses

\begin{equation}
\begin{split}
 m_\chi^2 = m_\phi^2+\frac{\lambda_\text{mix}v^2}{4},  \quad m_\eta^2 = m_\phi^2-\frac{\lambda_\text{mix}v^2}{4}.
\end{split}
\label{eq:phi mass splitting}
\end{equation}

\noindent This splitting can cause the lightest charged state's mass to become tachyonic, spontaneously breaking U$(1)_{EM}$ and giving the photon mass. This places a constraint that $m_{\phi}^{2} > \lambda_{\text{mix}} v^{2}/4$. 

The two remaining, neutral degrees of freedom are the real and imaginary parts of the third component of $\phi$, denoted $\phi_R^0$ and $\phi_I^0$, respectively. These states will be nearly degenerate with mass $m_{\phi}$ -- a very small splitting is generated which vanishes as $\kappa \to 0$.  Throughout we will use $\phi$ to refer to all of these states collectively and their individual names when specificity is required. %The phenomenology of these states differ in important ways and will be discussed in a benchmark scenario presented below.\\ 

\vspace{.1cm} \noindent \emph{Potential Minimization:}  The $\kappa$ term in the potential was introduced in order for the $\phi$ particles to decay, but also has other important effects that can constrain the model.  Once the electroweak and dark symmetries are broken, this term induces a vev for the real, neutral component of $\phi$. The size of this vev is

\begin{equation}
 \langle\phi\rangle = \frac{\kappa v^2 v_D}{4\sqrt{2}m_\phi^2}.
\end{equation}

\noindent Since this is only in the neutral component, U$(1)_{EM}$ remains unbroken, but it does shift the $W$ boson mass, with a contribution to the $T$ parameter 

\begin{equation}
 T_{\langle \phi \rangle}\sim  10^{-3}\, \kappa^2 \left(\frac{v_{D}}{1 \GeV}\right)^{2}\left(\frac{200 \text{ GeV}}{m_\phi}\right)^{4},
 \label{eq:T phi vev}
\end{equation}

\noindent which is very small as long as the dark photon scale is sub-GeV.  In addition, there is a one loop contribution to $T$ from the $\phi$ particles due to their mass splitting \cite{Lavoura:1993nq} which in the limit of small splitting goes as 
\begin{eqnarray}
T_\text{loop} \sim \frac{\lambda_\text{mix}^2 v^4}{192\pi  s_W^2 c_W^2 m_Z^2 m_\phi^2} \sim 0.1\, \lambda_\text{mix}^2 \left(\frac{200 \text{ GeV}}{m_\phi}\right)^{2}.
\label{eq:T mass splitting}
\end{eqnarray}
Contributions to $S$ are negligible, so to be consistent with electroweak precision constraints requires $T< 0.2 \text{ (95\% C.L.)}$ \cite{Agashe:2014kda}, putting a lower  bound on $m_\phi$ (from \eqnref{eq:T mass splitting}) and an upper bound on $\kappa$ (from \eqnref{eq:T phi vev}).\\

The $\kappa$ term also causes mixing between $\phi_R^0$, $h_{D}$, and the SM higgs.  This leads to a correction to the $\mu_D^2$ term of size $\kappa^2 v^4/(16 m_\phi^2)$. Thus, a large hierarchy between the dark and electroweak scales requires a tuning in the value of $\mu_{D}^{2}$.  The severity of this tuning depends on $\kappa$, and for certain regions of parameter space this tuning can be small. It is however interesting that the tuning in this model is indirectly observable. This is in contrast to the SM where the details of tuning depend on some unknown, as-of-yet-unobservable higher scale. If KM with $SU(2)_{L}$ is observed, this model will provide insight into the validity of tuning as a theoretical constraint.

\vspace{.1cm} \noindent \emph{Fixed Target Benchmark:}  Now lets consider a benchmark set of parameters, chosen in order to remain within the region of immediate interest to fixed-target experiments:  $m_{A_D}=0.1\GeV$ and $g_D=0.5$.  This choice implies that $v_D=0.2\GeV$, and we set $m_{h_D}=0.4\GeV$ so that the dark higgs can decay into two dark photons. Note that the dark higgs and photon masses are negligibly small relative to electroweak scale masses, so we can safely neglect them in later formulas.   We also set $\lambda_\text{mix} = 1$ which puts a lower limit on $m_\phi$ of 155\,GeV due to the electroweak precision constraint. In our analysis we specifically explore the range $150\GeV<m_\phi<500\GeV$ in order to be relevant for collider searches while remaining in the $10^{-5}<\epsilon<10^{-3}$ window, though it should be kept in mind that precision electroweak constraints exclude the small part of this region $m_{\phi}<155 \GeV$. Additionally, in this low mass region, the lightest charged state will have mass $m_{\eta} < 100 \GeV$ which is in tension with results from LEP searches for charged particles, e.g. \cite{Ackerstaff:1998si}.

\vspace{.1cm} \noindent \emph{Decays:} A $\phi$ particle can decay directly into gauge and higgs bosons through the $\kappa$ term, or undergo cascade decays through its mass states by radiating $W^{(*)}$ bosons.  The cascade decay rate, in the large $m_{\phi}$ and massless fermion limit, is
%%
%\begin{eqnarray}
%\Gamma(\chi^\pm\rightarrow W^{\pm *}\phi^0_{R,I}) &&=\Gamma(\phi^0_{R,I}\rightarrow W^{\mp *}\eta^\pm)\nonumber \\ \label{eq:cascade}
% && = \sum_{f\bar{f}'}\frac{N_c G_f^2 \Delta m^5}{15\pi^3}
%\end{eqnarray}
%%
\begin{align}
\Gamma(\chi^\pm\rightarrow W^{\pm *}\phi^0_{R,I})&=\Gamma(\phi^0_{R,I}\rightarrow W^{\mp *}\eta^\pm)\nonumber\\
&= \sum_{f\bar{f}'}\frac{N_c G_f^2 \Delta m^5}{15\pi^3}
\label{eq:cascade}
\end{align}

\noindent  where $G_f$ is the Fermi constant, $\Delta m$ is the mass splitting between $\phi$ states, and $f \bar{f}'$ includes all fermion pairs except the top-bottom pair for which the splitting $\Delta m$ is too small to produce.  The $\kappa$ mediated decay rates, in the limit that $m_{h_{D}},m_{A_{D}} \to 0$, are
\begin{eqnarray}
  \Gamma(\phi_R^0\rightarrow hh_D) &=& \Gamma(\phi_I^0\rightarrow hA_D) \nonumber\\
&=& \frac{\kappa^2v^2}{64\pi m_\phi^3}\left(m_\phi^2-m_h^2\right), \\
  \Gamma(\chi^\pm\rightarrow W^\pm h_D) &=& \Gamma(\chi^\pm\rightarrow W^\pm A_D) \nonumber \\
   &=& \frac{\kappa^2v^2}{128\pi m_\phi^4 m_\chi^3}\left(m_\chi^2-m_W^2\right)^3,\\
  \Gamma(\eta^\pm\rightarrow W^\pm h_D) &=& \Gamma(\eta^\pm\rightarrow W^\pm A_D) \label{eq:two body decay rates}  \\
   &=& \frac{\kappa^2v^2}{128\pi m_\phi^4 m_\eta^3 }\left(m_\eta^2-m_W^2\right)^3. \nonumber
\end{eqnarray}

The decay phenomenology depends sensitively on $\kappa$. If $\kappa$ is sufficiently small the cascade decays will dominate, and heavier $\phi$ will tend to decay down to the lightest state, $\eta^\pm$, emitting two fermions via an off-shell $W$ per step, followed by the $\eta^\pm$ decaying half the time to $W^\pm h_D$ and half the time to $W^\pm A_D$. On the other hand, if $\kappa$ is large, $\kappa$ mediated decays dominate with the neutral components of $\phi$ decaying as $\phi_R^0 \to h h_D, \phi_I^0 \to h A_D$ and $\eta^\pm,\chi^\pm$ decaying to $W^\pm h_D, W^\pm A_D$ equally.  Note that the simplicity of the decays are a consequence of our benchmark choice.  As the value of $v_D$ is increased from our benchmark, additional decay modes due to $\kappa$ become more important, {\it e.g.}~$\phi_R^0 \to hh, ZZ,WW$ and $\eta^\pm  \to W^\pm h, W^\pm Z$.  However, since these decay rates are proportional to $v_D^2$, only when $v_D \gtrsim 100 \GeV$ do these start to become important and thus in the intensity frontier parameter space we do not expect these decays to have appreciable rates.  

\begin{figure}[!htpb]
\begin{center}
\includegraphics[scale=.65]{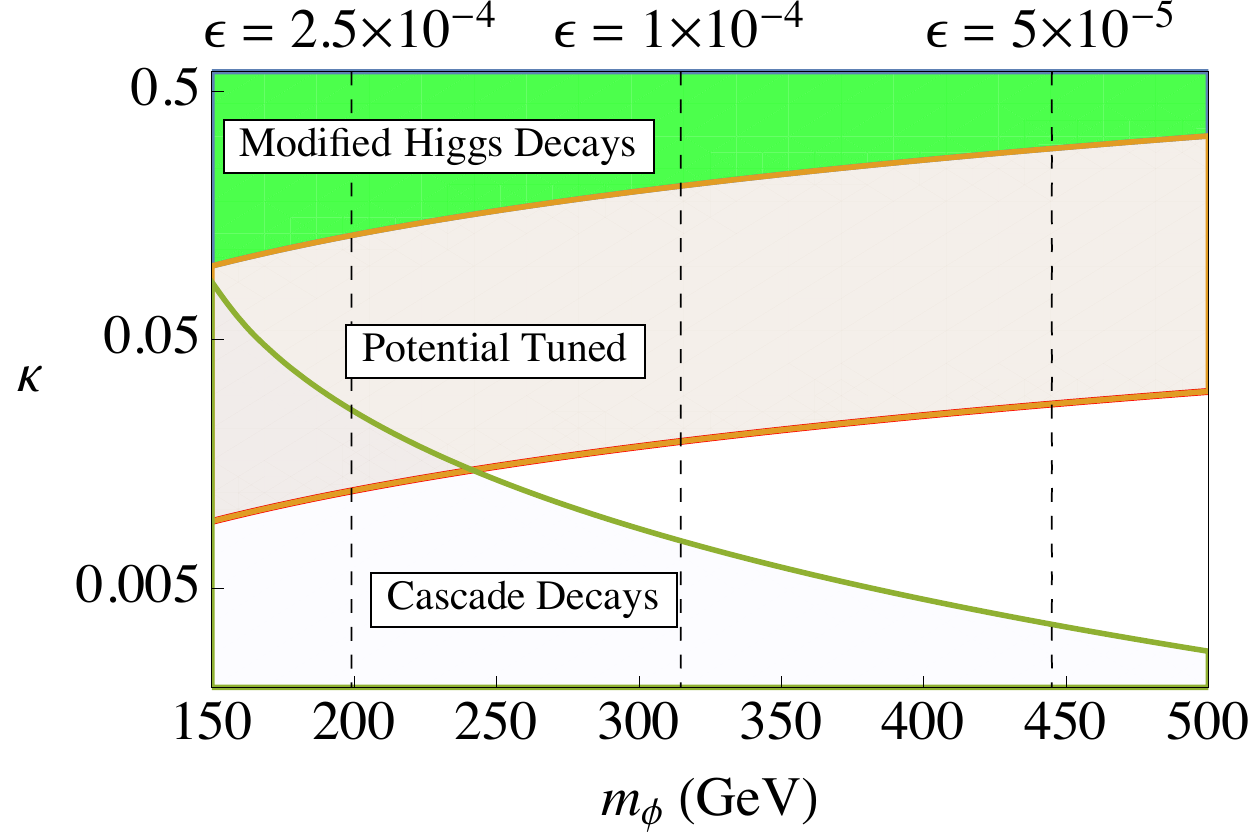}
\end{center}
\caption{This figure shows regions of interest in the ($m_{\phi}$,$\kappa$) plane. Starting from the top, the regions are where the new higgs decays are greater than $10\%$ of its SM expected total width (green region), where $\mu_{D}^{2}$ is tuned to $>10\%$ (above the thick orange curve) and where the electroweak cascade decays are faster than the $\kappa$ decays (below the green curve). Vertical dashed lines mark values of $\epsilon$, labeled at the top.}
\label{fig:kappa constraint}
\end{figure}

In \figref{fig:kappa constraint}, we highlight some of the important regions of our benchmark parameter space.  Some characteristic values of $\epsilon$ are given at three $m_\phi$ values in dashed lines, though these can be scaled up or down by changes in $g_D, \lambda_\text{mix}$.   The green line denotes the value of $\kappa$ where the cascade decays are comparable to the $\kappa$ induced decays below which the off-shell cascade decays dominate.  The middle region of \figref{fig:kappa constraint} shows where the tuning in $\mu^2_D$ is worse than $10\%$, and the last region at the top shows when the SM higgs has new decays with a branching ratio greater than 10\%, which will be discussed below. 
%Note that for $m_{\phi} < 200 \GeV$ decays of $m_{\eta}$ will no longer be in the Goldstone equivalence limit \cite{Cornwall:1974km,Vayonakis:1976vz} and rates will deviate from these formulae due to decays into transverse $W$ bosons.  %The analogous decay of $\chi_{\pm}$ is a bit slower because of its larger mass, but limits to the same value for $m_{\phi}^{2} \gg \lambda_\text{mix}v^2/4$.\\
 
\begin{figure}[!htpb]
\begin{center}
\includegraphics[scale=.94]{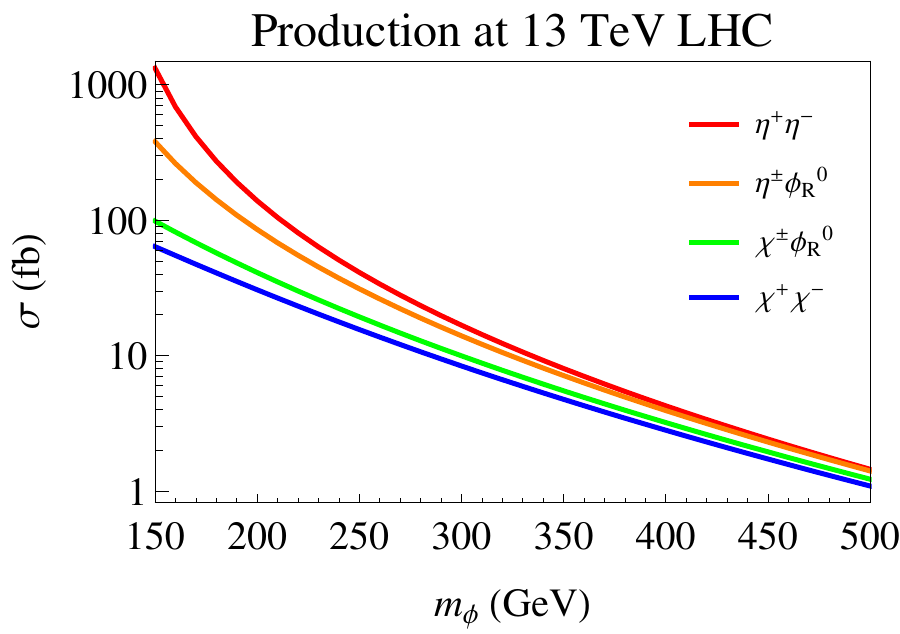}
\end{center}
\caption{This figure shows the pair production cross section for various mass states of $\phi$ at the $\sqrt{s} = 13\,$TeV LHC. The different curves are: two $\eta$ states (red), an $\eta$ and a $\phi^{0}_{R}$ or $\phi^{0}_{I}$ (orange) a $\chi$ and a $\phi^{0}_{R}$ or $\phi^{0}_{I}$ (green) and two $\chi$ states (blue).  The legend is arranged in order of decreasing cross section.  Curves were generated using the \texttt{cteq6l} parton distribution function of CalcHEP \cite{Belyaev:2012qa}.}
\label{fig:cross section}
\end{figure}

\vspace{.1cm} \noindent \emph{Production Rates:} In order to observe these decays, $\phi$ particles will need to be produced, which at a hadron collider proceeds predominantly through Drell-Yan production. The production cross sections at the 13\,TeV LHC are shown in \figref{fig:cross section}.  We used FeynRules \cite{Alloul:2013bka} to generate our Lagrangian and CalcHEP \cite{Belyaev:2012qa} to generate the events using the \texttt{cteq6l} parton distribution function for the proton. Pair production of the neutral particles does not occur due to the lack of photon, $Z$ couplings.  Also, production rates for $\phi_I^0$ are identical to $\phi_R^0$ and so are not included on the plot.   

The strategy for $\phi$ searches should start with adaptations to the existing searches for dark photons and lepton jets \cite{ArkaniHamed:2008qp, Aad:2012qua,Khachatryan:2015wka,Aad:2015sms}.  All events contain either $h_D$ or $A_D$ particles produced at significant boosts, which coupled with the decay $h_D \to A_D A_D$, will lead to many events with boosted lepton pairs. For small enough $\epsilon$, many of these $A_D$ decays will be displaced. If the value of $\kappa$ is small, where cascade decays dominate, there will also be soft leptons or jet activity from the off-shell $W$'s.  An interesting signal in this regime is the possibility of same sign $\eta$ production due to the cascade decays of $\phi_R^0,\phi_I^0$ going equally into $\eta^\pm$ (see \eqnref{eq:cascade}). Their subsequent decay produces a like-sign pair of $W$'s leading to same sign lepton events in addition to the lepton jets of the event.  On the other hand, if the value of $\kappa$ is large, there can be other associated objects like the SM higgs bosons produced in $\phi_R^0,\phi_I^0$ decays (see \eqnref{eq:two body decay rates}), which could be of interest in terms of tagging or reconstructing the events. To summarize, this scenario's predominant collider signal is lepton jets in association with  $W,h$ with mass resonances between a lepton jet and the $W$ or $h$.  

Since the benchmark's dark photon mass restricts it to electron decays, the lepton jets could be challenging to pick out.  Boosted electron pairs are much more difficult to distinguish from jets and in fact, most existing lepton jet searches rely on muons (with significant constraints only for  $M_{A_{D}} > 2 m_{\mu} \sim 0.2 \GeV$).  To overcome these challenges, some promising strategies could be to look for displaced jets and/or jets with significant electromagnetic energy deposit.  We leave studies of such issues as well as existing LHC constraints and discovery reach for such particles to future work.  

\vspace{.1cm} \noindent \emph{SM Higgs Phenomenology:} This model also predicts new decays for the SM higgs. The dominant new decays are into dark higgs bosons and dark gauge bosons. The kinetic mixing operator itself, \eqnref{eq:kin mix op}, generates new decays of the higgs to a dark photon and either a $Z$ or a photon.  The widths of these decays are
\begin{align}
  \Gamma(h&\rightarrow h_Dh_D) = \Gamma(h\rightarrow A_DA_D) =\frac{\kappa^4v^6}{512\pi m_hm_\phi^4}\\
  \Gamma(h&\rightarrow A_{D} \gamma) = v \frac{\epsilon^{2}}{8 \pi} \left(\frac{ m_{h}}{v}\right)^{3}\\
    \Gamma(h&\rightarrow A_{D} Z) \cong  \Gamma(h \rightarrow A_{D} \gamma) \times .2 \times \left(\frac{\, c_{W}}{s_{W}}\right)^{2} 
\label{eqn:higgsdecay}
\end{align}
\noindent again we take the limit where $A_{D}$ and $h_{D}$ are massless. These new decay widths are indirectly constrained by the relatively good fits of the SM higgs decay signal strengths \cite{ATLAS-CONF-2015-044}. As an approximation of this constraint, the top green region of \figref{fig:kappa constraint} shows where higgs decays into the dark sector exceed 10\% of the SM higgs total width.  In particular, decays of the higgs involving the dark photon are a direct consequence of the kinetic mixing term, and provide a model independent signal of nonabelian kinetic mixing. For $\epsilon \sim 10^{-3}$ the branching ratio of the higgs to a dark photon will be $Br(h \to A_{D} + Z/\gamma) \sim 4 \times 10^{-4}$. There is potential for the LHC to detect these higgs decays, if the dark photon is heavier than our benchmark. For example, if $m_{A_D} \sim 0.6-60$ GeV, the LHC can be sensitive to the dark photon through higgs decays into $2A_D$ \cite{Curtin:2014cca} and a recent LHC analysis constrains $Br(h\to 2A_D) \gtrsim 3\times 10^{-4}$ for $m_{A_D} = 15-60$ GeV \cite{Aad:2015sva}.  While the fixed target parameter space motivates searches at much lower dark photon masses, a simple modification of our benchmark can give these heavier masses.   In these modified benchmarks, if one improves the higgs branching ratio constraint to $BR_\text{new} < BR_\text{limit}$, this would constrain the range $\kappa > 0.25 (m_\phi/200 \GeV)Br_\text{limit}^{1/4}$.   As our formulas and discussion show, increasing $m_{A_D}$ to these larger values, either through increasing $g_D$ or $v_D$, changes very little in the $\phi$ phenomenology, however, in this heavier parameter space correlated signals at the intensity frontier could only be seen at future flavor factories for $m_{A_D}<10 \GeV$.
%\noindent \emph{Discussion:} The extrapolation of our benchmark scenario to other choices of the parameters is straightforward. The scaling of decays and $\epsilon$ are given by the expressions above, and the production of $\phi$ is only dependent on $m_\phi$ unless at low enough $m_{\phi}$ that $\eta^\pm$ becomes very light, which depends on $\lambda_\text{mix}$. For constant $m_{\phi}$ the value of $\kappa$ at which the dark higgs quadratic coupling becomes tuned depends only on the dark higgs mass and scales linearly with it so long as $m_{\phi} \gg m_{h_{D}}$.\\

\vspace{.1cm} \noindent \emph{Conclusions:} In this letter, we have argued for a direct connection between current intensity frontier searches for dark photons and the signals of new particles at the LHC.  The connection occurs if KM involves a nonabelian gauge symmetry, since the mixing operator requires higgs fields to be gauge invariant and thus closely ties the mediator particle mass to the vev of the higgs and the strength of KM.  To illustrate this, we wrote down a simple model where the only KM which occurs is between a new dark U$(1)$ gauge symmetry and $SU(2)_L$.  This requires a scalar triplet $\phi$ of $SU(2)_L$ which is charged under the dark U$(1)$.  Analyzing the model, we looked at the constraints and briefly considered the phenomenology of the $\phi$ particles at the LHC which could be searched through simple modifications of existing dark photon searches.

Aside from our simple model, there are obvious extensions to explore.  Fermionic mediators, mixing with a nonabelian dark gauge symmetry and incorporating dark matter are all intriguing modifications, which will all produce the same, model-independent correlation of signals.  Interestingly, these directions all tend to lead to larger multiplicity in the dark sector, suggesting that the model in this paper is unique in its simplicity.   Investigation of these directions, as well as a detailed collider study of this model is forthcoming.

%The ideal result for this scenario would be for a dark photon that kinetically mixes with the SM photon to be observed at a mass of $m_{A_D} \sim 100 \MeV$ with a KM strength of $\epsilon \sim 10^{-4}$, while at the same time discovering a new particle at the LHC with mass $m \sim 300 \GeV$ which decays primarily into SM gauge bosons and missing energy, perhaps associated with lepton jets.  Such a coincidence would provide extremely strong evidence for a model of this kind.\\

To conclude, KM of the SU$(2)_{L}$ of the SM and an abelian dark sector is timely and well motivated given the current run of the LHC, ongoing fixed target experiments, and potential next generation flavor factories.  The connection it draws between intensity and energy frontier experiments is unambiguous and leads to correlated signals at these experiments, promising unprecedented insight into the physics of the dark sector.

\vspace{.1cm} \noindent \emph{Acknowledgements:} The authors would like to thank J. Mardon for helpful discussions regarding the new decays of the SM higgs. This work was supported in part by the Department of Energy under grants DE-SC0009945 (G.B. and S.C.) and DE-SC0011640 (C.A.N.).

\bibliography{NAKineticMixing}

\end{document}